# Efficient $T_2$ mapping with Blip-up/down EPI and gSlider-SMS ($T_2$-BUDA-gSlider)

Xiaozhi Cao[1,2,3], Congyu Liao[2,3 *], Zijing Zhang[2,4], Siddharth Srinivasan Iyer[2,5], Zhifeng Chen[2,3,6], Wei-Ching Lo[7], Huafeng Liu[4], Kang Wang[8], Hongjian He[1], Kawin Setsompop[2,3,9], Jianhui Zhong[1,10] and Berkin Bilgic[2,3,9]

[1] Center for Brain Imaging Science and Technology, Department of Biomedical Engineering, Zhejiang University, Hangzhou, Zhejiang, China

[2] Athinoula A. Martinos Center for Biomedical Imaging, Massachusetts General Hospital, Charlestown, MA, USA

[3] Department of Radiology, Harvard Medical School, Charlestown, MA, USA

[4] State Key Laboratory of Modern Optical Instrumentation, College of Optical Science and Engineering, Zhejiang University, Hangzhou, Zhejiang, China

[5] Department of Electrical Engineering and Computer Science, Massachusetts Institute of Technology, Cambridge, MA, USA

[6] School of Biomedical Engineering, Guangdong Provincial Key Laboratory of Medical Image Processing, Southern Medical University, Guangzhou, China

[7] Siemens Medical Solutions, Boston, MA, USA

[8] Department of Neurology, the First Affiliated Hospital, School of Medicine, Zhejiang University, Hangzhou, China

[9] Harvard-MIT Department of Health Sciences and Technology, Cambridge, MA, USA

[10] Department of Imaging Sciences, University of Rochester, NY, USA

* Correspondence:

Congyu Liao, PhD, CLIAO2@mgh.harvard.edu, Athinoula A. Martinos Center for Biomedical Imaging, Building 149, Room 2301, 13th Street, Charlestown, MA, 02129 USA

## Abstract

**Purpose**: To rapidly obtain high isotropic-resolution $T_2$ maps with whole-brain coverage and high geometric fidelity.

**Methods**: A $T_2$ blip-up/down echo planar imaging (EPI) acquisition with generalized Slice-dithered enhanced resolution ($T_2$-BUDA-gSlider) is proposed. A radiofrequency (RF)-encoded multi-slab spin-echo EPI acquisition with multiple echo times (TEs) was developed to obtain high SNR efficiency with reduced repetition time (TR). This was combined with an interleaved 2-shot EPI acquisition using blip-up/down phase encoding. An estimated field map was incorporated into the joint multi-shot EPI reconstruction with a structured low rank constraint to achieve distortion-free and robust reconstruction for each slab without navigation. A Bloch simulated subspace model was integrated into gSlider reconstruction and utilized for $T_2$ quantification.

**Results**: In vivo results demonstrated that the $T_2$ values estimated by the proposed method were consistent with gold standard spin-echo acquisition. Compared to the reference 3D fast spin echo (FSE) images, distortion caused by off-resonance and eddy current effects were effectively mitigated.

**Conclusion**: BUDA-gSlider SE-EPI acquisition and gSlider-subspace joint reconstruction enabled distortion-free whole-brain $T_2$ mapping in 2 min at ~1 $mm^3$ isotropic resolution, which could bring significant benefits to related clinical and neuroscience applications.

## Introduction

High resolution $T_2$ mapping has shown great potential in clinical and neuroscience applications, including but not limited to epilepsy[1,2], glioma[3], monitoring tumor progression[4], brain maturation[5], femoroacetabular impingement[6] and Parkinson disease[7]. However, the use of conventional spin-echo acquisitions at multiple TEs[8], which is considered the gold standard approach for $T_2$ mapping, require lengthy acquisition time (e.g. about 60 min for 10 TEs acquired at $1{\times}1{\times}5$ $mm^3$ resolution with limited slice coverage), which prohibits its clinical application. Other approaches including DESPOT$_2$[9], MR fingerprinting[10,11], 3D turbo spin-echo (TSE) $T_2$ mapping[12,13], multi-echo acquisition[14] and model-based acceleration[15–17] have allowed for faster $T_2$ parameter estimation. These techniques can reduce the scan time and improve clinical feasibility, but still require several minutes to provide whole-brain high-resolution $T_2$ maps, for example, 17 min for DESPOT$_2$[9], 11-16 min for model-based $T_2$ mapping[18] and 5-8 min for 3D MR fingerprinting (while also provides $T_1$ mapping)[11,19].

Another potential solution is using spin-echo (SE) echo planar imaging (EPI) to reduce the scan time. To achieve whole-brain coverage, 2D SE-EPI efficiently utilizes the idle time during the lengthy TR required for $T_1$ recovery to acquire data from multiple slices. While this enables faster scans, it is limited to thick slice acquisition as 2D encoding fails to provide sufficient SNR for high resolution quantitative imaging. Recently, a slab encoding technique named generalized slice-dithered enhanced resolution (gSlider)[20] was proposed, which utilizes RF pulses with different excitation profiles to encode individual thin slabs. The gSlider technique fully utilizes the long TR by acquiring data from all slabs sequentially while enjoying higher SNR due to its volumetric encoding, which is crucial for whole-brain, high-resolution $T_2$ mapping.

Despite their high efficiency k-space encoding, drawbacks common to all EPI-based readout strategies are $T_2$- or $T_2$*-related voxel blurring, geometric distortion and voxel pile-ups stemming from $B_0$ inhomogeneity. These preclude high in-plane resolution imaging with a single-shot EPI readout. A common practice is to use parallel imaging techniques[21–24] to reduce the effective echo spacing, but these are usually limited to $R_{in\text{-}plane} \leq 4$ in-plane acceleration using modern receive arrays. Multi-shot EPI approaches segment and acquire a plane of k-space data across multiple TRs, thereby reducing the geometric distortion and blurring and permitting higher in-plane resolution imaging. Shot-to-shot phase variations due to physiological noise in multi-shot EPI need to be accounted for to enable successful image reconstruction. A navigator echo can help estimate these variations[25], but this comes at the cost of reduced acquisition efficiency since navigation further prolongs the acquisition time. Navigator-free methods such as Hankel structured low-rank constrained parallel imaging techniques[26–30] have been introduced to address this problem. However, such advanced reconstruction techniques require a larger number of shots at very high in-plane acceleration (e.g. 4-shots acquisition with $R_{in\text{-}plane} = 8$[30]) to achieve good image quality and low distortion, which again reduces the acquisition efficiency.

To mitigate the geometric distortion, FSL topup[31,32] is commonly used, where two EPI acquisitions with reverse phase-encoding polarities, one with blip-up and another with blip-down, are acquired separately to estimate a field map and perform distortion correction. In this way, distortion-free images can be obtained with only 2-shots of data, which helps reduce the acquisition time compared to standard multi-shot EPI using only blip-up encoding. Based on this approach, a hybrid-space SENSE method[33] was proposed to jointly reconstruct the blip-up/down shots with estimated field maps and incorporated the phase differences into the forward model. This method was able to reduce the g-factor penalty and improve the SNR over single-shot EPI. However, since the hybrid-space SENSE requires explicit knowledge of phase variations between the blip-up and -down shots, their inaccurate estimation may introduce reconstruction artifacts and noise amplification[34].

In this work, we propose to combine gSlider encoding and blip-up/down acquisition (BUDA) to achieve high-resolution and distortion-free $T_2$ mapping with whole-brain coverage. First, we incorporate Hankel structured low-rank constraint into our BUDA reconstruction to recover distortion-free images from blip-up/down shots without navigation. To utilize the similarity among different TEs, we introduce a model-based gSlider-subspace joint reconstruction to recover high-resolution thin-slice images. This projects time-domain images into temporal coefficient maps with a temporal basis[35,36] and then the reconstructed coefficients are used to obtain quantitative $T_2$ maps. The proposed method enables distortion-free high-quality whole-brain $T_2$ mapping with ~1 mm isotropic resolution in 2 minutes. This work is an extension of our earlier work which was reported in abstract form[37].

## Method

### BUDA-gSlider acquisition

Figure 1A shows the sequence diagram of proposed multi-slab BUDA-gSlider acquisition. Ten different TEs are acquired to obtain different $T_2$ weighted contrasts. Successive TE volumes are acquired with alternating blip-up or -down polarity. E.g. $TE_1$ data are encoded with 5 gSlider shots using blip-up polarity, whereas the 5 RF encoding shots in the following $TE_2$ acquisition use blip-down phase encoding. The entire acquisition thus requires a total of 50 shots ($N_{TE}\times N_{RF}$ = 10×5, where $N_{TE}$ is the number of different TEs used for generating different $T_2$ weightings, and $N_{RF}$ is the number of RF pulses used for encoding one slab).

To obtain whole-brain coverage efficiently, we utilize the idle time of each TR to acquire data from other slabs. With the 5-mm slab thickness, acquiring 26 slabs will correspond to a 130 mm FOV in the slice direction, thus providing the desired whole-brain coverage. In this way, the entire acquisition time could be described as $T_{acq}=TR\times N_{TE}\times N_{RF}$. To further accelerate the acquisition, we also introduced blipped-CAIPI encoding for simultaneous multi-slab acquisition[38] to reduce the TR to 2400 ms with multi-band (MB) factor 2. With this, we have obtained $T_{acq}$=2.4×10×5=120 sec.

For gSlider slab-encodings, the 90° excitation pulses are designed to achieve a highly independent encoding. The five 90° excitation pulses are named as $RF_1$ to $RF_5$ sequentially and used to encode the same slab, which is 5 times as thick as the desired slice thickness. In this work, the targeted slice resolution is 1 mm, and the corresponding thickness of each individual slab is 5 mm. Since each of these RF pulses excites a slab rather than a single slice, there is volumetric SNR gain for each acquisition[20].

### k/t BUDA reconstruction

With acquired interleaved blip-up and blip-down shots, distortion-free images of each RF encoding can be jointly reconstructed using the pipeline shown in Figure 1C, which includes the following:

i. The blip-up EPI data (yellow lines in Figure 1B) and the blip-down EPI data (red lines in Figure 1B) were separately reconstructed using SENSE to obtain five pairs of images with ten different TEs, SENSE blip-up and SENSE blip-down shown in Figure 1C. The SENSE blip-up/down images have different geometric distortion due to opposite phase-encoding direction as the arrow indicates.
ii. All of the ten SENSE blip-up/down images were jointly processed in FSL topup (http://fsl.fmrib.ox.ac.uk/fsl) to estimate field maps [31]. The advantage by using five pairs of SENSE blip-up/down images for field estimation over one single pair will be shown later in Results section.
iii. The estimated field maps were incorporated into the Hankel structured low-rank constrained joint reconstruction for two contiguous blip-up/down data [34]. This can be described as:

$$min_x \sum_{t=2n-1}^{t=2n} \left\| \mathrm{F}_t \mathrm{E}_t \mathrm{C} b_{t,r} - d_{t,r} \right\|_2^2 + \|\mathcal{H}(b)\|_* \qquad (1)$$

Where *t* is the index of TE and *r* is the index of RF encoding, and *n* runs from 1 to 5, representing the index of blip-up/down pair. $\mathrm{F_t}$ is the undersampled Fourier operator in $t^{th}$ TE's shot, $\mathrm{E_t}$ is the estimated off-resonance information, $\mathrm{C}$ are the ESPIRiT [23] coil sensitivity maps estimated from distortion-free gradient-echo calibration data, $b_{t,r}$ is the distortion-free image and $d_{t,r}$ are the $t^{th}$ TE's and $r^{th}$ RF-encoding k-space data. The constraint $\|\mathcal{H}(b)\|_*$ enforces low-rank prior on the block-Hankel representation of the blip-up and blip-down data $b_{t\&t+1,r}$, which is implemented by consecutively selecting 9×9 neighborhood points in k-space from each shot and then concatenating them in the column dimension. This reconstruction is implemented by using a projection onto convex sets (POCS) like[39,40] iteration with the tolerance of 0.01% RMSE between two successive iterations. Please note that the reconstructed images $b_{t,r}$ are still RF-encoded slab images. After using this approach to reconstruct five RF encodings (*r*=1,2…5), a total of 50 distortion-free k/t BUDA images were obtained.

### gSlider-subspace joint reconstruction

A model-based gSlider-subspace joint reconstruction was proposed and shown in Figure 2:

i. A dictionary comprising signal evolution curves of $T_2$ from 1 to 1000 ms was built. With principle component analysis, the first ten principle components were selected as the temporal basis **Φ**, shown in Figure 2A. With this basis, the desired high-resolution slab images $x_{t,s}$ (where *s*=1,2…5 is the slice index within a RF-encoded slab) could be expressed as $\mathbf{\Phi} c_{t,s}$, where $c_{t,s}$ is the temporal coefficient map.

ii. To jointly reconstruct all the 50 shots of different TEs and RF encodings, all the BUDA reconstructed images $b_{t,r}$(*t*=1,2…10, *r*=1,2…5) were expanded in the column dimension as $b_{\text{exp}}$, resulting in a 50×1 vector for each voxel with contributions from all TEs and RF encodings. To meet the form of $b_{\text{exp}}$, $c_{t,s}$ is also expanded but along the TE-Slice dimension as $c_{\text{exp}}$.

iii. Figure 2B shows the gSlider-subspace joint reconstruction, which can also be described as:

$$min_{c_{\text{exp}}} \left\| \mathbf{A}_{\text{exp}} \mathbf{\Phi}_{\text{exp}} c_{\text{exp}} - b_{\text{exp}} \right\|_2^2 + \lambda_{\text{Tik}} \left\| c_{\text{exp}} \right\|_2^2 \qquad (2)$$

where $\mathbf{A}_{\text{exp}}$, a matrix with size of 50×50, is the expanded RF-encoding matrix which includes ten RF-encoding matrices $A$(This matrix contains the excitation profiles of all RF-encoding pulses and can be calculated using Bloch simulation). $\mathbf{\Phi}_{\text{exp}}$ is the expanded temporal basis including five temporal basis **Φ** corresponding to all the five RF encodings. $\|.\|_2^2$ denotes Tikhonov regularization and $\lambda_{\text{Tik}}$ is the gSlider Tikhonov subspace regularization parameter and was set to 0.3 to achieve a high SNR gain and sharp partition resolution based on a retrospective experiment, which will be shown in Results section. In addition, the noise-like weak coefficient components (e.g. $c_{4-10,s}$ for five slices from one same slab) were removed for the purpose of denoising and improving the conditioning of the reconstruction. The supplementary Figure S1 shows exemplar ten coefficient maps from a single slice, where the first three coefficient components capture ~95% of whole signal intensity while the rest three weak coefficient maps are noise-like.

iv. The temporal coefficient maps were projected back to time domain to recover the high-resolution thin-slice images with different $T_2$ contrasts by $\mathbf{\Phi}_{\text{exp}}^T c_{\text{exp}}$. Then, the recovered images were used to get the $T_2$ maps using template matching with the pre-calculated dictionary voxel-by-voxel.

**In-vivo validation**

To validate our proposed method, one 28-years-old male healthy volunteer was scanned with the approval of Institutional Review Board.

The proposed $T_2$-BUDA-gSlider data were collected using the following protocol: $R_{in\text{-}plane}$=4, partial Fourier 6/8, multi-band factor=2, TR=2400 ms, 10 different TEs (from 49ms to 121ms with a gap of 8ms, where the 49ms is the shortest possible TE that EPI readout could achieve and 121ms is selected to cover the $T_2$ range of common tissues in brain). A phase-encoding shift $\Delta k_y$ of 2 was set between the blip-up and blip-down shots to improve the k-space coverage and the parallel imaging reconstruction. 26 thin slabs (slab thickness=5mm) were acquired with 5 RF encodings for each slab, resulting in ~1-mm slice thickness (influenced by the selection of $\lambda_{\text{Tik}}$, which will be discussed later in Result section) for the final high-resolution thin-slice images. With 1×1 $mm^2$ in-plane resolution, the resulting voxel size was 1×1×~1 $mm^3$ with FOV=220×220×130 $mm^3$. For each TE and RF encoding, both blip-up and -down shots were acquired as reference, resulting in a total of 100 shots and the total acquisition time of 240s ($T_{acq} = TR \times N_{TE} \times N_{RF\text{-}encoding} \times$blip-up/down = 2.4s×10×5×2). Then it was subsampled by alternatively selecting blip-up or blip-down shots in the TE dimension to 50 shots, corresponding to an acquisition time of 120s. A FOV-matched low-resolution 2D gradient-echo (GRE) was also acquired to obtain distortion-free sensitivity maps for BUDA reconstruction.

To test the accuracy of the proposed method, multi-TE single-echo spin-echo (SE) data were also collected as the gold standard to estimate $T_2$ values. The protocol of SE sequence was set the same as the proposed method but with a slice thickness of 5 mm. $T_2$ values were then fitted voxel-by-voxel using a non-linear least square method to obtain the $T_2$ maps. In addition, to demonstrate the distortion-free property of the proposed BUDA-gSlider, a 3D fast spin-echo was also acquired as reference. To compare the $T_2$ values in specific subcortical regions by using proposed method and gold standard SE, a $T_1$-weighted MPRAGE sequence was acquired as the input for Freeserfer's subcortical segmentation[41].

All studies were performed on a 3 Tesla (T) MAGNETOM Prisma scanner (Siemens Healthcare, Erlangen, Germany) with a 64-channel head receiver coil. Computations were performed on a Linux (Red Hat Enterprise) server (with Core i7 Intel Xeon 2.8 GHz CPUs and 64GB RAM) using MATLAB R2014a (The MathWorks, Inc., Natick, MA).

## Results

Figure 3 shows the RF-encoded slab images of RF1/TE1 (i.e. 49 ms) by using hybrid-space SENSE and BUDA reconstruction, respectively. As indicated by the blue arrows, the individual SENSE reconstructed images for blip-up and blip-down shots exhibited significant geometric distortions compared to the 3D FSE reference, while the results from BUDA are consistent with the reference. As indicated by the red arrows, the BUDA results had reduced noise amplification and artifacts compared to hybrid-space SENSE. Figure 3B shows the RF-encoded slab images of different TEs using BUDA reconstruction, which provide the different $T_2$ contrasts for subsequent $T_2$ mapping.

Figure 4 shows $T_2$ maps obtained by using different $\lambda_{\text{Tik}}$ values for Tikhonov regularization of gSlider-subspace joint reconstruction. The impulse responses of the central slice within an RF-encoded slab  were shown in the bottom. Here we replaced $b_{\text{exp}}$ with a gSlider RF-encoded delta function of the central slice and pass it through the proposed method to obtain the impulse response. The reference results were obtained by using the fully-sampled 100-shots acquisition with $\lambda_{\text{Tik}}$ of 0.1. Compared with the reference, when $\lambda_{\text{Tik}}$was selected between 0.1 to 0.3, a high SNR gain and sharp partition resolution could be achieved. For $\lambda_{\text{Tik}}$ of 1, significant blurring in partition direction could be observed.

Figure 5A shows four exemplar slices of $T_2$ maps from the gold standard SE (top) and the proposed method with $\lambda_{\text{Tik}}$=0.3 (middle). To directly compare the gold standard method and the proposed method regardless of resolution difference, $T_2$ maps averaged across 5 adjacent slices from BUDA-gSlider were also shown in the bottom. $T_2$ values from six specific regions (including white matter, gray matter, pallidum, thalamus, caudate and putamen) were evaluated and shown in Figure 5B for both methods. Estimated $T_2$ values were close to the gold standard SE but with faster acquisition (2 min vs 60 min), higher resolution ($1\times1\times1$ mm$^3$ vs $1\times1\times5$ mm$^3$), and larger slice coverage (130 mm vs 75 mm) which demonstrate the utility of $T_2$-BUDA-gSlider.

Supplementary Figure S2 shows the field maps (first column) estimated by using only one single pair of blip-up/down SENSE reconstruction images (top) and all five pairs of blip-up/down SENSE images (bottom), respectively. With more pairs of blip-up/down SENSE images as input for FSL topup, the filed map tends to become less noisy and more reasonable, and thus results in better quality for BUDA reconstruction and subsequent $T_2$ mapping.

## Discussion and Conclusion

The proposed $T_2$-BUDA-gSlider provides a new approach for rapid and distortion-free $T_2$ mapping with the ability to provide whole-brain coverage at ~1-mm$^3$ isotropic resolution in 2min, with an SNR-efficient RF-encoded SE-EPI acquisition. Compared with current state-of-the-art methods listed in the Introduction section, the proposed method is fast with good image quality. It takes advantages of the following:

i. Distortion-free EPI: By incorporating $B_0$-correction in the forward model, BUDA acquisition and its corresponding joint reconstruction can provide high-quality distortion-free images while the EPI readout can enable fast acquisition.

ii. SNR gain of volumetric gSlider acquisition: Since the individual SENSE blip-up and -down reconstructions are RF-encoded slab images, they benefit from the volumetric SNR gain of using thicker slabs as opposed to conventional 2D acquisitions. The estimated $T_2$ maps also benefit from the SNR boost.

iii. Improved image quality with gSlider-subspace joint reconstruction: We utilized the similarity of images with different TEs to generate better quality $T_2$ maps through joint reconstruction. This has further capitalized on the SNR gain from gSlider and BUDA techniques, which has improved the image quality of the $T_2$ maps. In addition, we used the temporal basis to project images of all TEs and RF encodings into a joint reconstruction model and then reduced the noise by eliminating the weak coefficient components.

The $T_2$ values measured by the proposed method were in accordance with the gold standard SE. We also demonstrated the distortion-free property of the proposed method by comparing the images with 3D FSE data, where BUDA was seen to mitigate the distortion caused by off-resonance and eddy current effects. By comparing the field map estimated by single pair and multiple pairs (i.e. five) of SENSE blip-up/down images, it could also be found that the latter one enabled better reconstruction results for both BUDA and $T_2$ mapping.

Since the Tikhonov regularization used in the gSlider-subspace joint reconstruction would introduce side lobe, the partition resolution could be influenced especially when $\lambda_{\mathrm{Tik}}$ is big. Therefore the $\lambda_{\mathrm{Tik}}$ should be selected within the range from 0.1 to 0.3, corresponding to a total side-lobe proportion from 7.08% to 16.64%. Thus a sharp partition resolution as well as high SNR gain could be achieved. And $\lambda_{\mathrm{Tik}}$=0.3 was selected in Figure 5 based on the consideration of better image quality.

One limitation is the relatively narrow range of acquired TEs (i.e. from 49ms to 121ms, where the 49ms is the shortest possible TE based on current acquisition parameters). We have employed these values to target gray and white matter in the brain, but this range needs to be adjusted to target short $T_2$ species (e.g. cartilage and myelin) or to better capture very long $T_2$ species (e.g. CSF).

Based on the BUDA-gSlider SE-EPI acquisition strategy and gSlider-subspace reconstruction, a rapid, distortion-free, high-resolution, whole-brain $T_2$ mapping approach is proposed. The proposed method could obtain whole-brain distortion-free $T_2$ maps with 1-$mm^3$ isotropic resolution in 2 min.

## Acknowledgement

This work was supported by:
National Institute of Biomedical Imaging and Bioengineering, Grant/Award Number: P41 EB030006, R01 EB019437, R01 EB020613, U01 EB025162 and R01 EB028797;
National Institute of Mental Health, Grant/Award Number: R01 MH116173;
NVIDIA GPU grant;
Zijing Zhang is supported by the China Scholarship Council for a 2-year study at Massachusetts General Hospital.

# Figure Caption

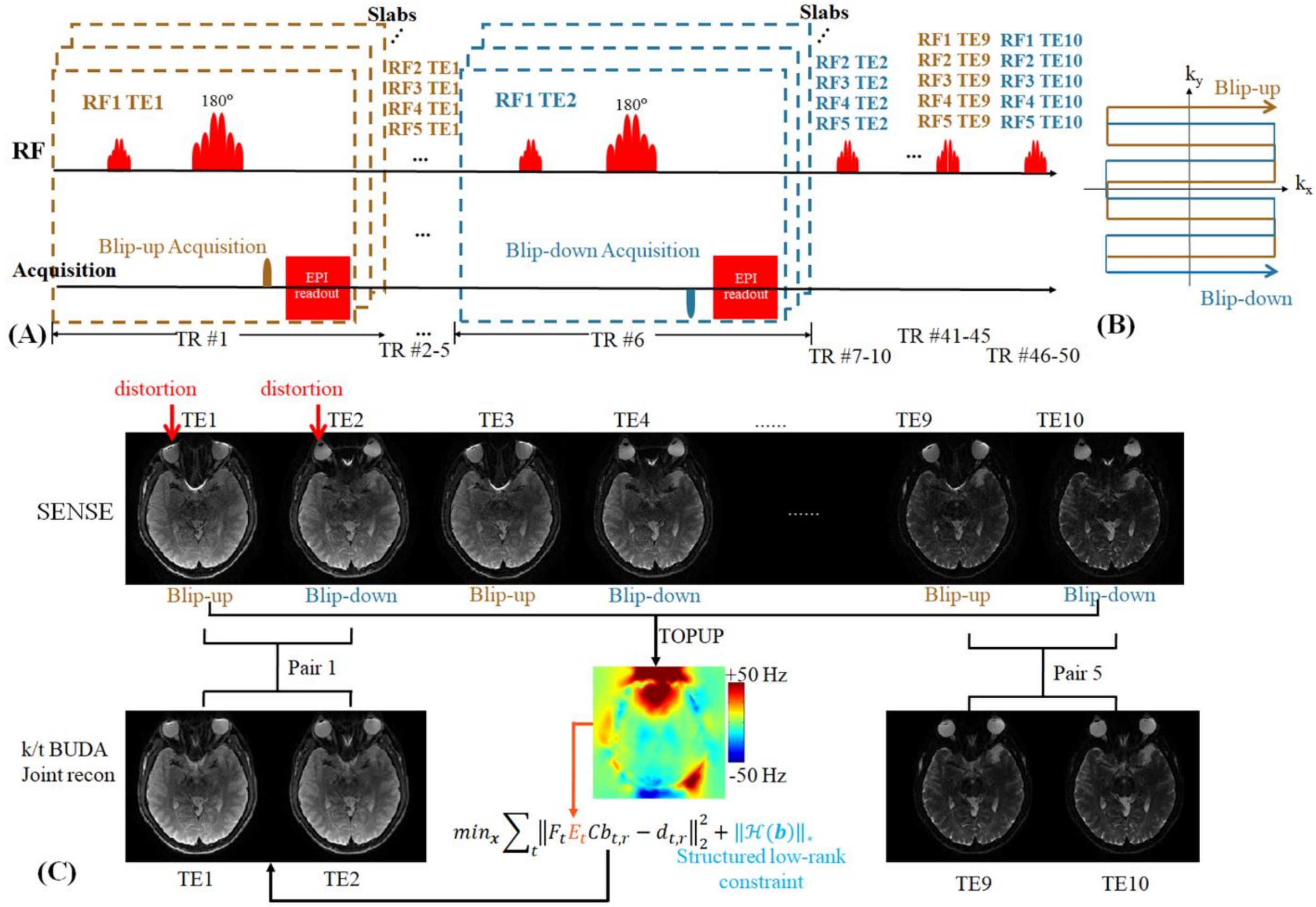

Figure 1

(A) The sequence diagram of the BUDA-gSlider acquisition with multi-TE and spin-echo EPI readout.

(B) The acquisition trajectory of the blip-up/down 2-shots EPI.

(C) The flowchart of k/t BUDA reconstruction. The reconstructed SENSE images by respectively using blip-up and -down shots are jointly used to estimate the field map using FSL topup. Then the field map is incorporated into the Hankel structured low-rank constrained forward model to jointly reconstruct distortion-free images.

## (A) Dictionary generation with subspace

signal intensity

0.5

50

100

TE(ms)

SVD

Φ1
Φ2
Φ3
Φ4
Φ5
Φ6
Φ7
Φ8
Φ9
Φ10

BUDA recon images @ 10 TEs

$\Phi^{-1}x = c$

$c_1$ $c_2(5\times)$ $c_3(10\times)$

signal evolution of $T_2$ values: [1:1000] ms

temporal basis $\mathbf{\Phi}$ containing 10 components

three coefficients $c$

## (B) gSlider-subspace joint reconstruction

RF-encoding matrix

temporal basis Φ

Coefficient maps

RF-encoded slab images

$A_1$ … $A_{10}$

Φ Φ Φ Φ Φ

$C_{1,1}$, $C_{2,1}$, $C_{3,1}$, …, $C_{10,5}$

$b_{1,1}$, $b_{2,1}$, $b_{3,1}$, …, $b_{10,5}$

$$\mathbf{A}_{exp} \times \mathbf{\Phi}_{exp} \times c_{exp} = b_{exp}$$

## (C) Template match

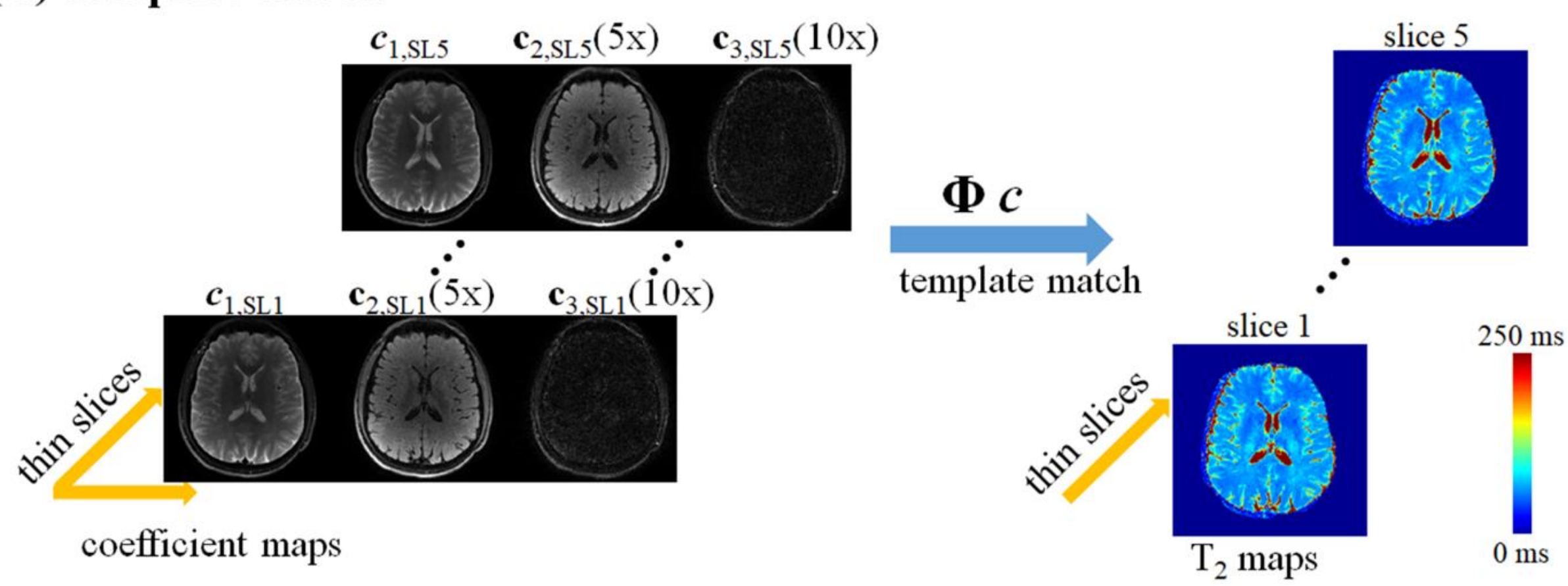

Figure 2

The gSlider-subspace reconstruction process including:

(A) Generating $T_2$ dictionary by using EPG algorithm and corresponding temporal basis by principle component analysis.

(B) The gSlider-subspace joint reconstruction which is used to generate the high-resolution thin-slice images by utilizing the similarity of images between different TEs and RF encodings.

(C) Template match the high-resolution thin-slice images with a pre-calculated $T_2$ dictionary to obtain the final $T_2$ maps.

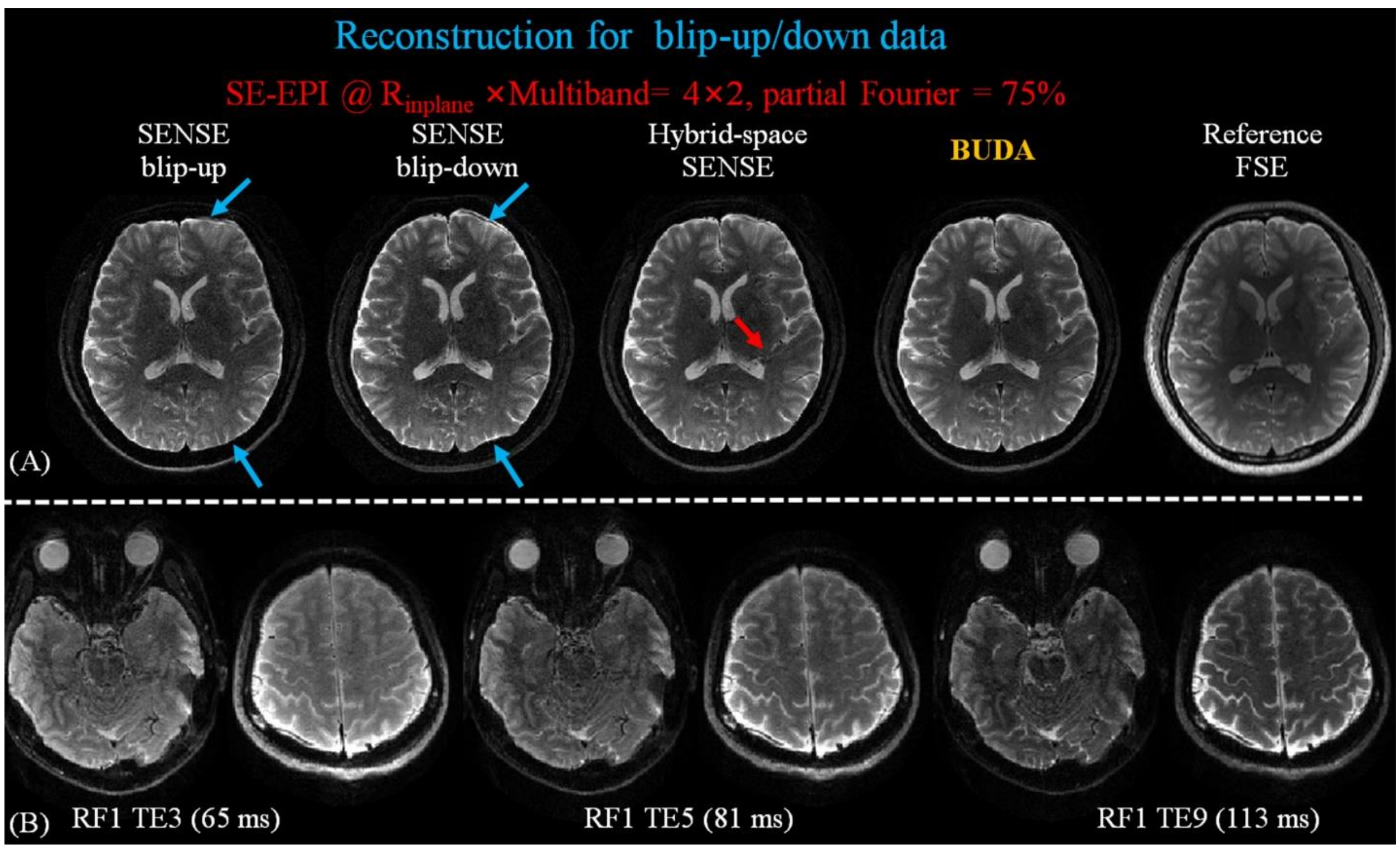

Figure 3

(A) Reconstructed RF-encoded slab images of RF1/TE1 and reference 3D-FSE images. The images using BUDA reconstruction has reduced artifacts compared to hybrid-space SENSE and high geometric fidelity consistent to the reference images.

(B) RF-encoded slab images of different TEs by using BUDA reconstruction.

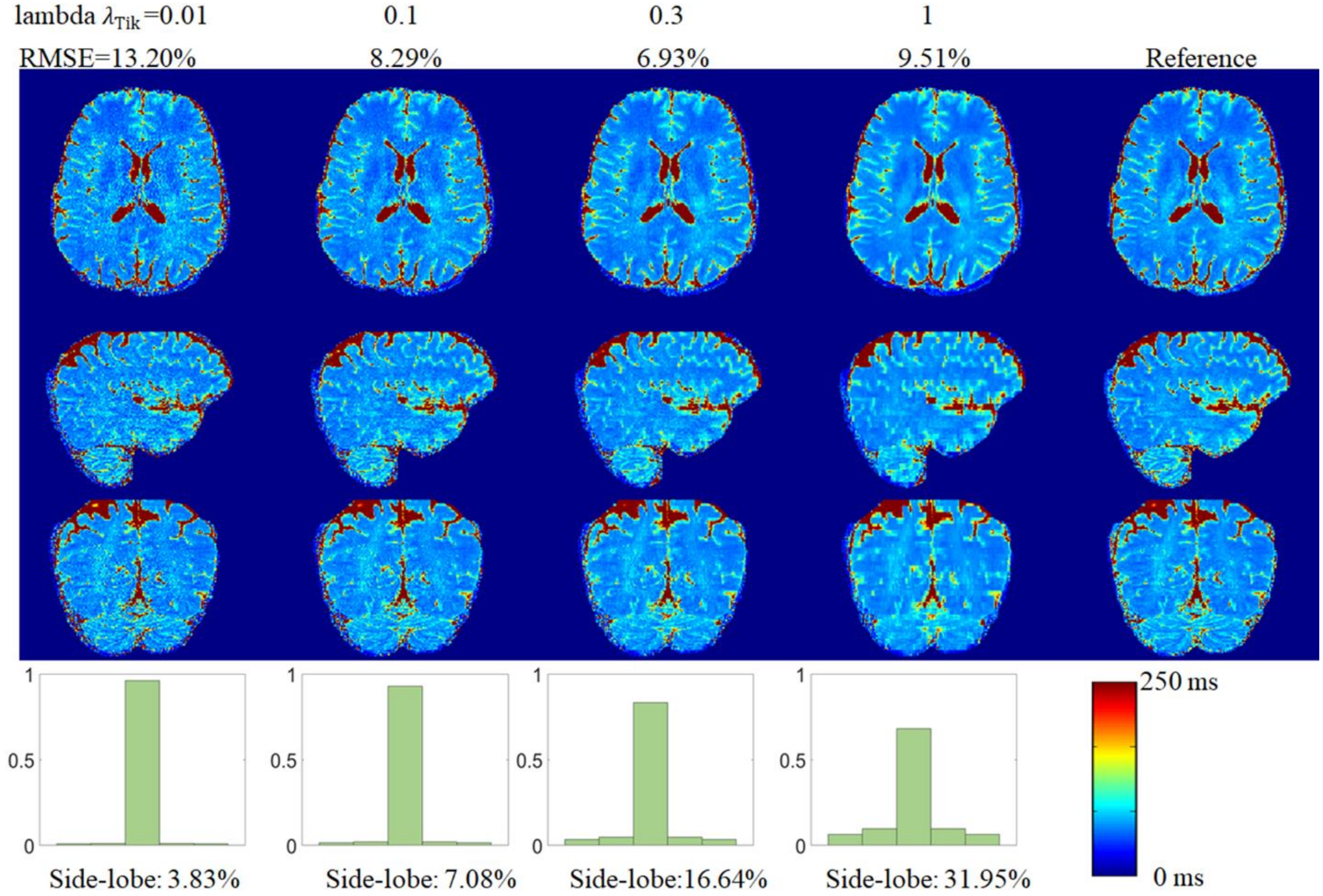

Figure 4

The comparison of T2 maps based on different lambda for Tikhonov regularization from 0.01 to 1. The reference is from the full-sampled 100-shots acquisition.

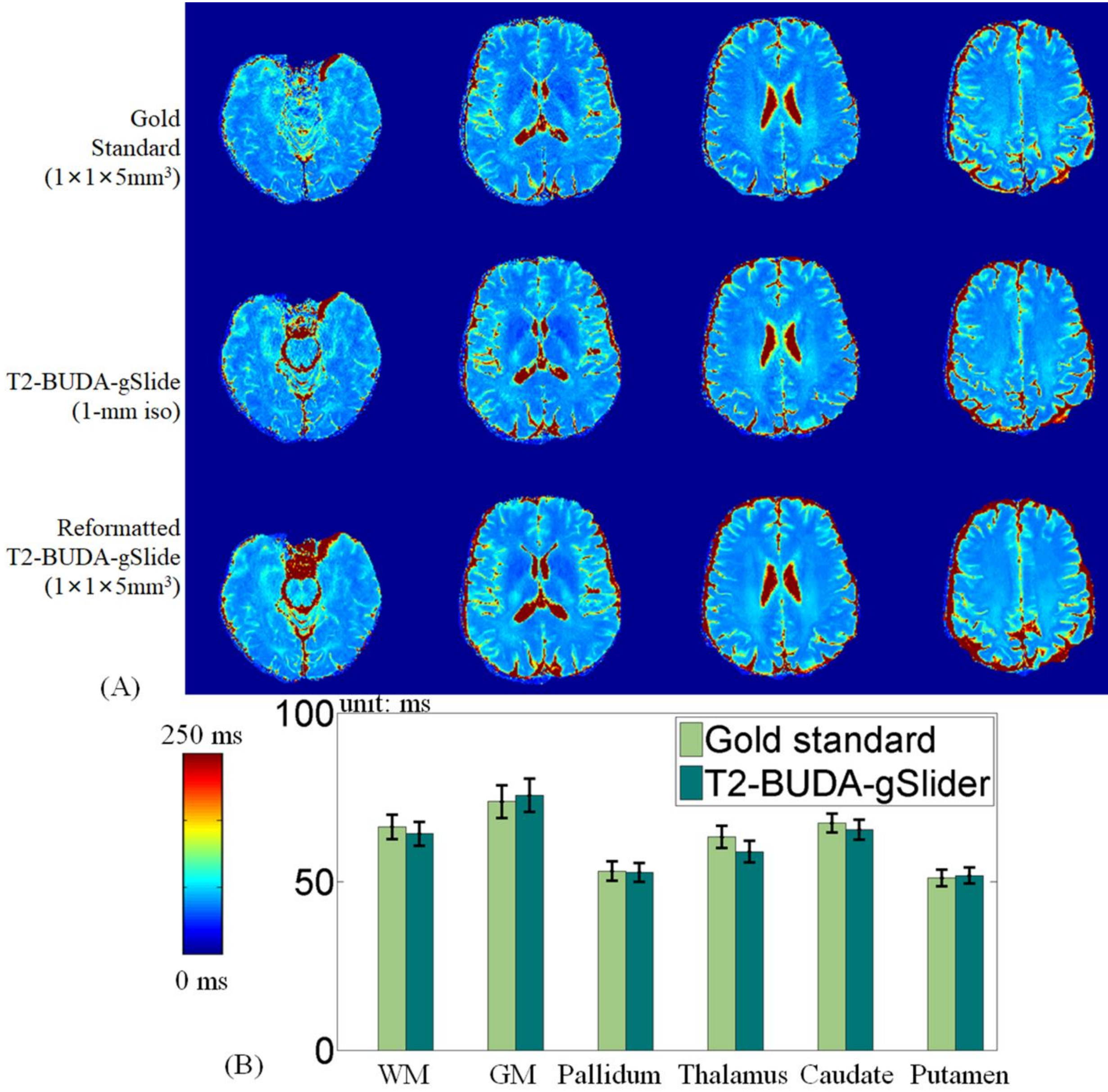

Figure 5

The comparison between the proposed $T_2$-BUDA-gSlider method and gold standard spin-echo method. The proposed method can achieve higher resolution (1×1×1 $mm^3$ vs 1×1×5 $mm^3$) and wider coverage (130 mm vs 75 mm in the slice direction) within short acquisition time (2 min vs 60 min).

# Supplementary Materials

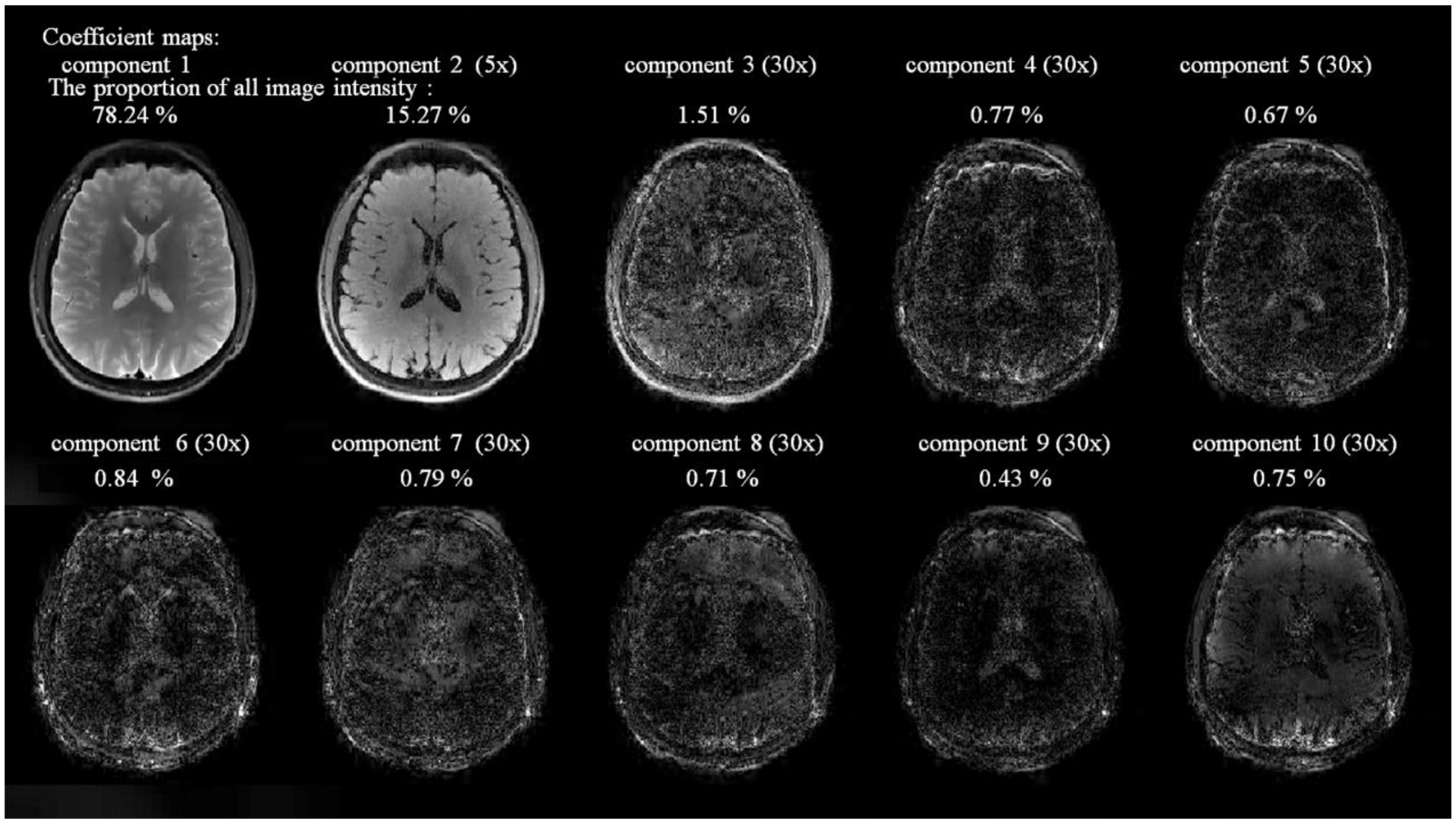

Supplementary Figure S1

The coefficient maps projected from the high-resolution thin-slice images by using a temporal basis.

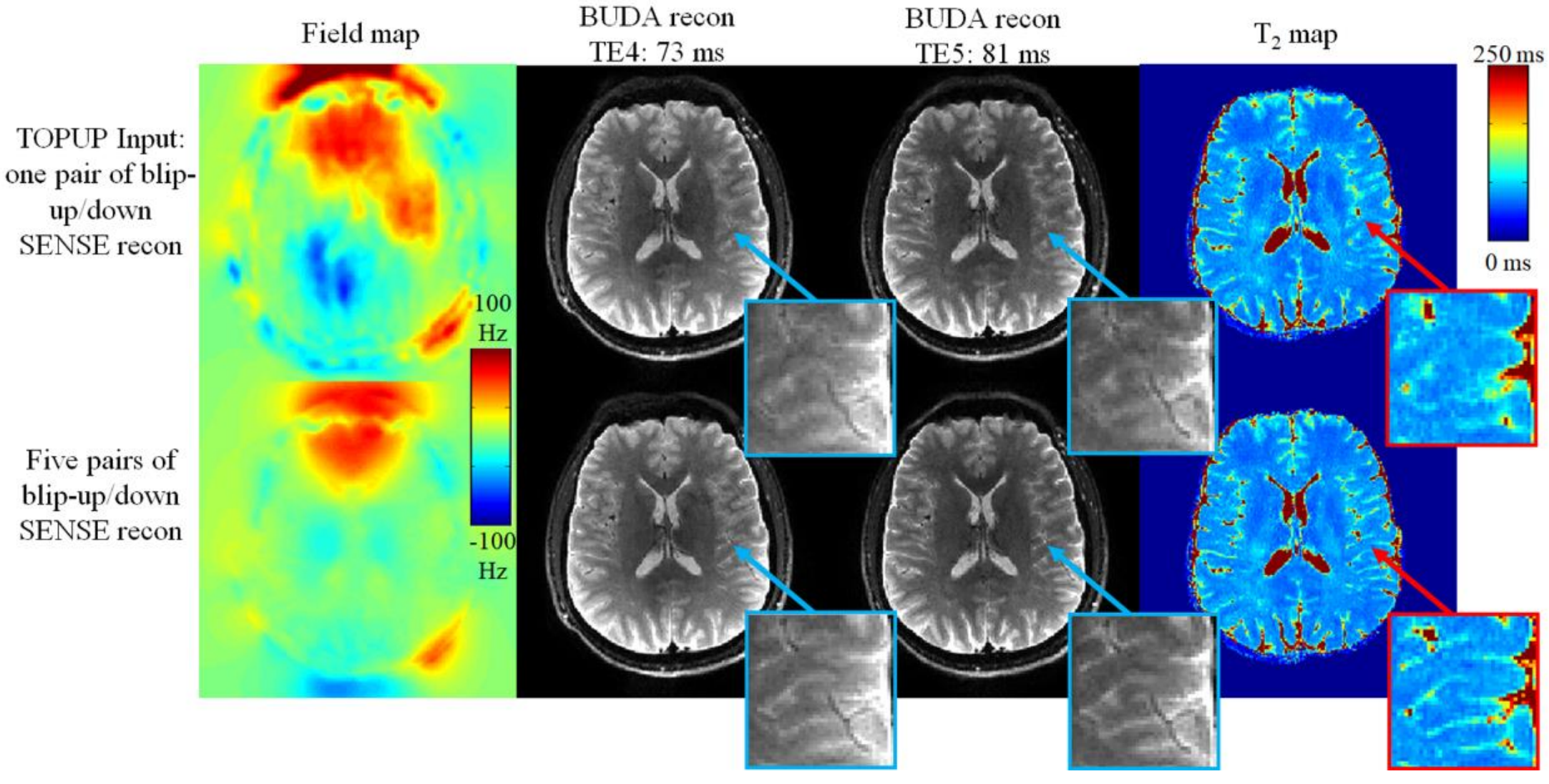

Supplementary Figure S2

Comparison between using one single pair of SENSE blip-up/down images to do field map estimation with using five pairs. The corresponding BUDA reconstruction results and $T_2$ maps are also included.